\documentclass[twocolumn,prd,superscriptaddress,showpacs,nofootinbib,floatfix,10pt]{revtex4-2}
\usepackage{epsfig}
\usepackage{graphicx}
\usepackage{epstopdf}
\usepackage{float}
\usepackage{subfigure}
\usepackage{diagbox}
\usepackage{tabstackengine}
\usepackage{amsmath,fourier,amssymb,physics,bm,slashed,tensor,ytableau}
\usepackage{soul}
\usepackage{tikz}
\usepackage{tikz-3dplot}
\tdplotsetmaincoords{80}{45}
\tdplotsetrotatedcoords{-90}{180}{-90}
\usepackage[eps]{pstricks}
\usepackage[normalem]{ulem}  

\newcommand{\pc}{$P_{c}$}
\newcommand{\br}{\mathcal{B}}
\newcommand{\pcftot}{$P_{c}(4312)$}
\newcommand{\jpsi}{$J /\psi$}
\newcommand{\gj}{$g_{J/\psi}$}

\newcommand{\sqrts}{$\sqrt{s}$}
\renewcommand\sout{\bgroup \color{red} \ULdepth=-.5ex \ULset}
\newcommand{\comment}[1]{}

\begin{document}

\title{The Electron-Ion Collider as A Prospective Facility for Pentaquark Measurements}
\author{In Woo Park}%
\affiliation{Department of Physics and Institute of Physics and Applied Physics, Yonsei University, Seoul 03722, Korea}

\author{Sungtae Cho}
\affiliation{Division of Science Education, Kangwon National
University, Chuncheon 24341, Korea}
\affiliation{Center for Extreme Nuclear Matters (CENuM), Korea University, Seoul, 02841, Korea}

\author{Yongsun Kim}%
\email{yongsun.kim@cern.ch}
\affiliation{Center for Extreme Nuclear Matters (CENuM), Korea University, Seoul, 02841, Korea}
\affiliation{Department of Physics, Sejong University, Seoul, 05006, Korea}
\affiliation{Brookhaven National Laboratory, Upton, 11973, USA}

\author{Su Houng Lee}%
\email{suhoung@yonsei.ac.kr}
\affiliation{Department of Physics and Institute of Physics and Applied Physics, Yonsei University, Seoul 03722, Korea}


\begin{abstract}
The Electron-Ion Collider provides a groundbreaking opportunity to study heavy pentaquarks with unprecedented precision, leveraging its high collision energy and beam spin polarization capabilities. As a representative case, we analyze electroproduction cross sections of \pcftot~under different spin-parity hypotheses using the vector meson dominance model. To ensure a parameter-free approach and minimize ambiguity, we incorporate results from the LHCb and GlueX experiments. To characterize the spin and the parity of \pcftot, we propose measuring the beam spin asymmetry and decay kinematic polarization, quantities that can be accurately determined by the ePIC detector.  Our approach can be extended to investigate other heavy pentaquarks produced in electron-proton and electron-deuteron collisions, as well as to study their interactions with nuclear matter in  electron-heavy ion collisions. We strongly encourage the EIC community to explore this potential and integrate pentaquark studies as a critical element of the scientific mission.

\end{abstract}

\maketitle


{\it Introduction:} The Electron-Ion Collider (EIC), currently under construction, is the groundbreaking experiment in nuclear and particle physics~\cite{eic-yellow}. It will facilitate electron-proton ($e+p$) and electron-nucleus ($e+$A) collisions, achieving unprecedented luminosity at center-of-mass (c.m.) energies of up to 141 GeV and 90 GeV per electron-nucleon pair, respectively. As a state-of-the-art facility, the EIC is designed to explore the inner structure of protons and nuclei with unmatched precision, shedding light on the fundamental mechanisms of quantum chromodynamics (QCD) and the origins of hadronic mass. Despite its ambitious objectives, the experimental program has overlooked an exciting opportunity: the study of exotic hadrons, particularly pentaquark states.

The LHCb's discovery of several heavy pentaquark states, $P_c$ ($uudc\bar{c}$)~\cite{LHCb:2015yax,LHCb:2019kea} and $P_{cs}$ ($uusc\bar{c}$)~\cite{LHCb:2020jpq}, marks a major milestone in exotic hadron research. However, their quantum properties, such as spin and parity, remain unknown due to the current constraints in experiments. In high-energy proton-proton ($p+p$) collisions, pentaquarks are typically produced within jet fragmentation processes and/or as feed-down products from heavier b-quark hadrons such as $\Lambda_b^0$ and $\Xi_b^-$. However, the small cross section of these states in hadronic processes, the difficulties in isolating these events in the trigger, and the substantial background noise in $p+p$ collisions impose significant limitations on precise measurements of their properties.  Another limitation of this channel is that it cannot produce pentaquarks heavier than $\Lambda_b^0$ and $\Xi_b^-$~— for example the yet-to-be-discovered $P_b~(uudb\bar{b})$ states.

In this letter, we propose that high-energy $e+p$ experiments at the EIC offer the optimal solution  for studying heavy pentaquarks. The minimal collision system that allows direct pentaquark production is $e+p$, as the total baryon number is one, enabling the EIC to produce pentaquarks with minimal background. Leveraging the high-energy photons radiated from polarized electrons, the EIC provides a unique platform to produce and study heavy pentaquarks. This is because the virtual photon can couple to heavy quark-antiquark pairs through vector meson dominance (VMD)\cite{Sakurai:1960ju}. Furthermore, the facility's ability to exploit electron polarization opens new possibilities for studying the spin and parity properties of these states.

We argue that the EIC is the ideal machine for precision measurements of pentaquarks. As an illustrative example, we explore the electroproduction cross sections of  \pcftot~ under various spin-parity hypotheses, specifically $J^P = \left(\frac{1}{2}\right)^{\pm}$, $\left(\frac{3}{2}\right)^{\pm}$, and $\left(\frac{5}{2}\right)^{\pm}$. This investigation underscores the EIC's potential to provide definitive insights into the quantum states of pentaquarks, representing a significant advancement in our understanding of exotic hadrons.

{\it Calculations:} 
Depending on the spin-parity assignment of \pcftot, its interactions with the proton and the $J/\psi$  are given by \cite{Shklyar:2004ba,Oh:2007jd,Shklyar:2004dy}:
\begin{align}\label{eq:jppclagrangian}
\mathcal{L}_{\text{int}}&=\begin{cases}\frac{g_{JpP_{c}}}{m_{J/\psi}}\bar{\psi}_{p}\begin{pmatrix}\sigma^{\mu\nu}\\\gamma_{5}\sigma^{\mu\nu}\end{pmatrix}F^J_{\mu\nu}\psi_{P_c}&~~ J^{P}=\left(\frac{1}{2}\right)^\pm,\\\\\frac{g_{JpP_{c}}}{m_{J/\psi}}\bar{\psi}_{p}\begin{pmatrix}\gamma_{5}\gamma^{\nu}\\\gamma^{\nu}\end{pmatrix}F^J_{\mu\nu}\psi^{\mu}_{P_c}&~~ J^{P}=\left(\frac{3}{2}\right)^\pm,\\\\\frac{g_{JpP_{c}}}{m_{J/\psi}^{2}}\bar{\psi}_{p}\begin{pmatrix}\gamma^{\nu}\\\gamma_{5}\gamma^{\nu}\end{pmatrix}\partial_{\alpha}F^{J}_{\mu\nu}\psi_{P_{c}}^{\mu\alpha}&~~J^{P}=\left(\frac{5}{2}\right)^\pm.
\end{cases}
\end{align}
Here, $\psi_{p}$ is the proton spinor, while  $\psi_{P_c}$, $\psi_{P_c}^{\mu}$ and $\psi_{P_c}^{\mu\alpha}$
are the spinors of the \pc~state with $J=\frac{1}{2}$, $J=\frac{3}{2}$ and  $J=\frac{5}{2}$, respectively. $F^J_{\mu\nu}=\partial_{\mu}A^J_\nu-\partial_\nu A^J_\mu$ where $A^{J}_{\mu}$ is the massive vector field.
Then, given the total decay width of \pcftot, \( \Gamma_{\text{tot}} = 9.8~\text{MeV} \), as measured by the LHCb collaboration \cite{LHCb:2019kea}, the coupling constant \( g_{JpP_c} \) can be calculated using the following partial decay width.
\begin{align}
\label{eq:couplingstrength}
\Gamma_{P_c\to p+J/\psi} & =\Gamma_{\text{tot}} \cdot \br 
=\frac{1}{8\pi}\frac{\abs{\bm{p_{f}}}} {m_{P_{c}}^{2}}\abs{\mathcal{M}}^{2}. 
\end{align}
$\mathcal{M}$ represents the matrix element obtained from Eq.~(\ref{eq:jppclagrangian}) for the decay process $P_c \to p + J/\psi$. $\bm{p_f}$ denotes the momentum of the produced proton in the rest frame of \pcftot~and 
$\br$ is the branching ratio for the $P_c \to p + J/\psi$ channel.  
Therefore, the width measurement from LHCb data imposes a constraint on the $g_{JpP_c}/\sqrt{\br}$~value.

\begin{figure}
\centering
\subfigure[]
{\includegraphics[scale=.72]{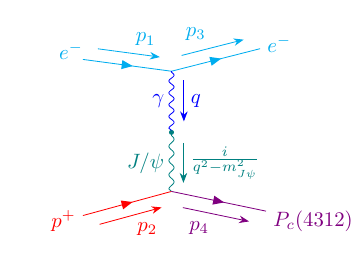}\label{fig:electroproduction}}
\subfigure[]{\includegraphics[scale=.68]{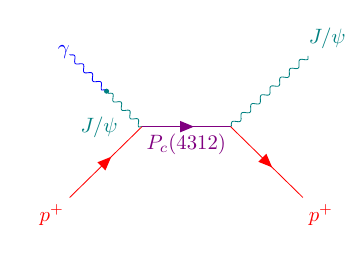}\label{fig:photoproduction}}
\caption{(a) Electroproduction channel to be measured by EIC.   (b) Photoproduction channel investigated by GlueX experiment~\cite{HillerBlin:2016odx}.}
\end{figure}

In $e+p$ collisions, the proton interacts with the electron via a photon as shown in Figure~\ref{fig:electroproduction}.
Under the VMD hypothesis~\cite{Klingl:1996by}, the emitted photon can be expressed as a combination of neutral vector meson currents. 
For the vector current composed of a $c\bar{c}$ pair, its coupling to the $J/\psi$, defined as $\mel{0}{J_{\text{em}}^{\mu}}{J/\psi} = -\frac{m_{J/\psi}^2}{g_{J/\psi}} \varepsilon_{J/\psi}^\mu$, can be determined through the $J/\psi \rightarrow e^+e^-$ decay, yielding  \gj~= 11.2. Such prescription effectively introduces the following form factor for the process in Figure~\ref{fig:electroproduction}.\begin{align}\label{eq:ggammapc}
g_{\gamma pP_{c}} = -\frac{e}{g_{J/\psi}} \frac{m_{J/\psi}^{2}}
{q^{2} - m_{J/\psi}^{2}} g_{JpP_{c}}
\end{align}
where $q$ denotes the four-momentum of the photon.

$\br$, the only free parameter in our model thus far, can be determined using the method outlined in Ref.~\cite{GlueX:2019mkq, JPAC:2016odx,HillerBlin:2016odx}. These studies extracted the upper limit of $\br$ for $J^{P} = \frac{3}{2}^{-}$ by ensuring that GlueX's \jpsi~photoproduction cross section in $\gamma + p$ collisions is properly described. This analysis resulted in $\br = 2.9\%$.  Using the values of $g_{\gamma pP_{c}}$ and $\br$, we calculated the cross section for the photoproduction of \pcftot~by employing the relativistic Breit-Wigner formula and setting $q^2 = 0$ in Eq.~(\ref{eq:ggammapc}) to extend the VMD for real photon propagators.

\begin{align}
\sigma_{\gamma p\to P_c\to J/\psi p} = & \frac{4\pi}{ \bm{p_i}^2}\frac{2J+1}{2\cdot2}\frac{s \Gamma_{\text{tot}}}{(s-m_{P_c}^2)^2+s\Gamma_{P_c}^2} \cdot \nonumber \\  
&  \Gamma_{P_c\to \gamma p}(s)   \cdot \br .
\end{align}
Here, $\abs{\bm{p_i}}=\frac{1}{2\sqrt{s}}\left(s-m_p^2\right)$ and $s=m_p^2+2m_{p}E_{\gamma}$, where $m_p$ is the proton mass and $E_{\gamma}$ is the photon energy in the Lab frame. 
We verified that our calculation is consistent with the GlueX result when we used $\br = 2.9\%$ for $J^{P} =\frac{3}{2}^{-}$ as suggested in~\cite{GlueX:2019mkq}.  Based on this validation, $\br$ values for other spin-parity cases are fixed so that the photoproduction cross section at the resonance peak energy ($\mathrm{E}_{\gamma} = 9.44~\mathrm{GeV}$) matches the result for $J^{P} = \frac{3}{2}^-$,  thus ensuring the compatibility between hypothesis and experimental data.  In the end, with the constraint on $g_{JpP_c}/\sqrt{\br}$, the values of $g_{JpP_c}$ and $\br$ can be obtained as summarized in Table~\ref{table:coupbr}, where other parameters used in the calculation are taken from PDG~\cite{Zyla:2020zbs}.

\begin{table}[H]
\centering
 \begin{tabular}{c c c c c c c}
 \hline\hline $J^P$ & $(1/2)^+$ & $(1/2)^-$ & $(3/2)^+$ & $(3/2)^-$ & $(5/2)^+$ & $(5/2)^-$ \\\hline $g_{JpP_c}$ & 0.048 & 0.032  & 0.16  & 0.10 & 0.45 & 0.72\\\hline BR[\%] & 1.6 & 3.7 & 1.2  & 2.9 & 0.84 & 0.32\\\hline\hline
\end{tabular}
\caption{ The dimensionless coupling constant and branching ratio for $P_c \rightarrow p+J/\psi$ for six spin-parity cases. }
\label{table:coupbr}
\end{table}

{\it Total cross section:} With all necessary ingredients in place, we can now compute the differential cross section for $e + p \to e + P_c$ at the EIC, which is given by the following formula:
\begin{align}
\left(\dv{\sigma}{\theta}\right)_{\text{c.m.}}&=\frac{2\pi\sin\theta}{64\pi^{2}s}\frac{\abs{\bm{p_f}}}{\abs{\bm{p_i}}}\abs{\mathcal{M}}^2,
\end{align}
where $\bm{p_i}$ and $\bm{p_f}$ denote the momenta of the incoming and outgoing particles, respectively, and $\theta$ represents the polar angle between the beam axis and \pc~in the c.m. frame.
The matrix element $\mathcal{M}$ involving spin $\frac{5}{2}$ can be calculated using the procedures given in \cite{Shklyar:2004ba,Oh:2007jd,Shklyar:2004dy}.

In favor of the ePIC detector~\cite{epic} in EIC, the differential cross section was calculated by transforming into the lab frame and as a function of pseudorapidity, $\eta = \ln\tan\frac{\theta_{\text{Lab}}}{2}$. The choice of $\eta$ is driven by its influence on the detector efficiency, which is largely dependent on this variable because of the cylindrical symmetry of the detector.  According to the convention of the EIC community, $\eta$ is positive in the proton-moving direction (forward) and negative in the electron-moving direction (backward).  
The electron and proton beam energies are set at 275 GeV and 18 GeV, respectively, representing the maximum energy configuration of the EIC and corresponding to a c.m. energy ($\sqrt{s}$) of 141 GeV. 
The cross section, $\dd[]\sigma / \dd[]\eta$, is evaluated within $\abs{\eta} \leq 4$, matching the coverage of the ePIC detector.  Subdetector components such as $\mu$RWell trackers, AC-LGAD trackers, dRICH detectors, and electromagnetic calorimeters are expected to reconstruct electrons and protons with nearly 100\% efficiency in this range. 

\begin{figure}
\centering
\includegraphics[scale=0.178]{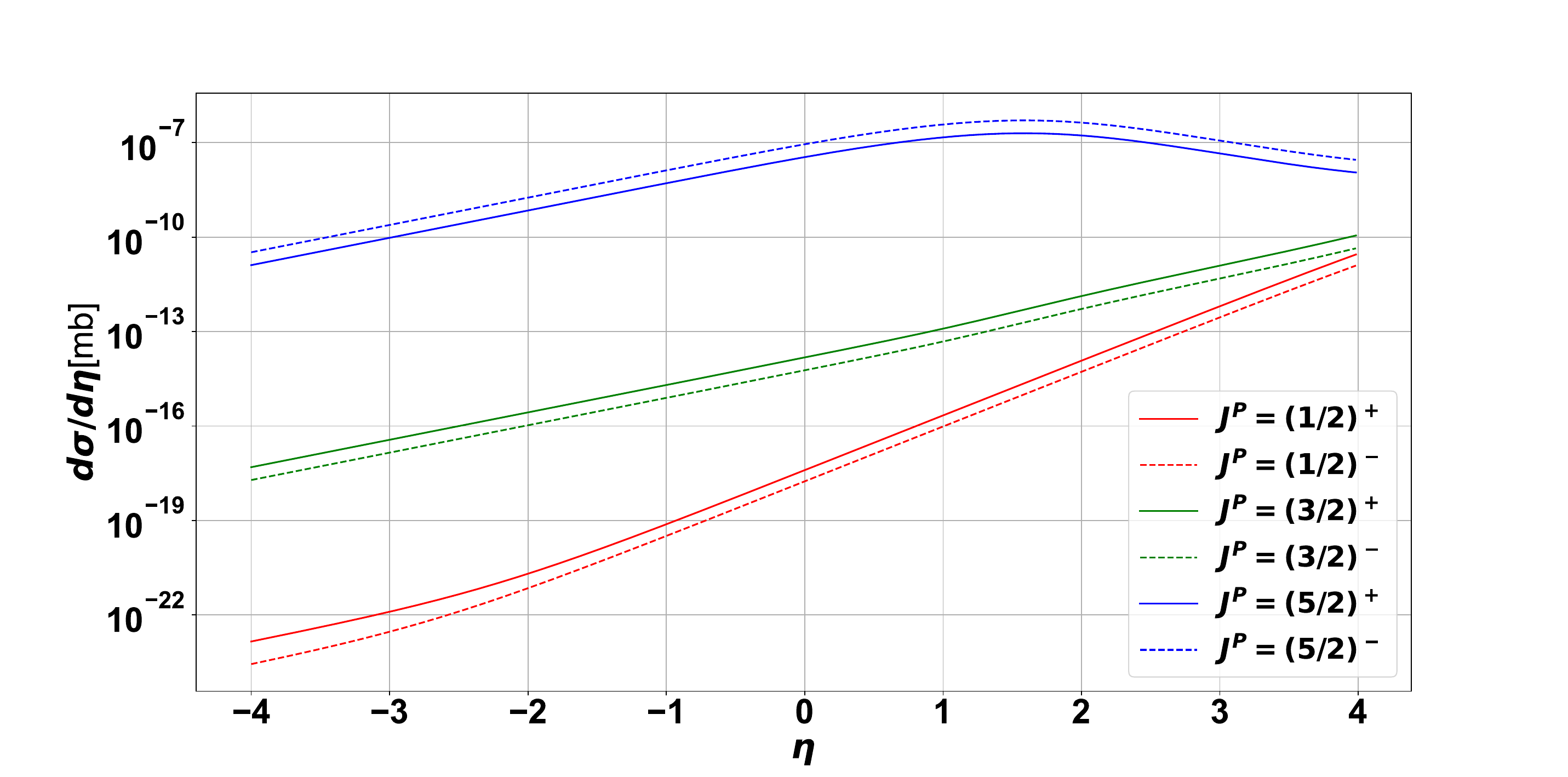}
\caption{Differential cross section of \pcftot~production in $e+p$ collisions at \sqrts=141 GeV for six spin-parity assignments.}\label{fig:unpolarized_eta}
\end{figure}

Figure~\ref{fig:unpolarized_eta} shows the differential cross sections for unpolarized beams. The cross section is larger for higher spin states. Such behavior is related to the larger couplings for higher spin states needed to reproduce the same partial decay width in Eq.~\eqref{eq:couplingstrength}, as these states have a smaller decay phase space near the threshold.
Another notable observation for $J = \frac{5}{2}$ is that the cross section of the negative-parity state is larger than that of the positive-parity state, in contrast to the behavior observed for $J =\frac{1}{2}$ and $J =\frac{3}{2}$.
This difference can be attributed to angular momentum.
Taking into account of the lower part of Figure~\ref{fig:electroproduction}, where the photon interacts with the proton and produces the \pc, conservation of parity and angular momentum leads to the possible choices of $l$ as given in Table~\ref{table:roam} for each spin-parity scenario. When the spin of \pc~is $\frac{1}{2}$ or $\frac{3}{2}$, the positive parity state has a larger relative orbital angular momentum, whereas for \pc~with spin-$\frac{5}{2}$, the negative parity possesses the larger orbital angular momentum. Larger $l$ results in a higher power of momentum, leading to a larger cross section at high energies.

\begin{figure}
\centering
\includegraphics[scale=0.175]{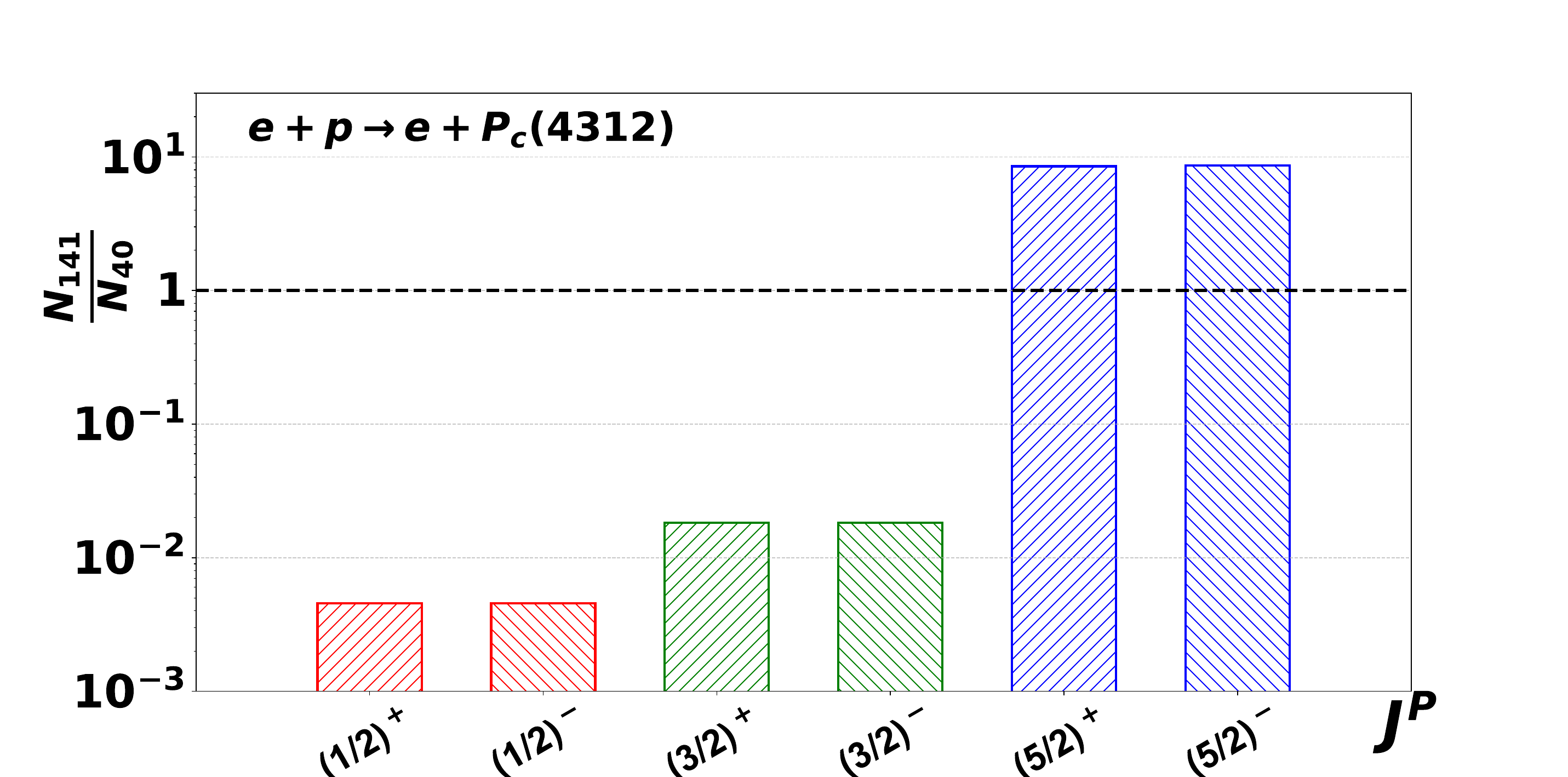}
\caption{Ratio of the integrated \pcftot~yields in $e+p$ collisions at \sqrts=141 GeV and \sqrts=40 GeV for six spin-parity assignments.  The pseudorapidity range is restricted to |$\eta$|<4 to account for detector geometry.} \label{fig:pcamount}
\end{figure}

\begin{table}[htbp]
\caption{The possible values of relative angular momentum $l$ for each spin-parity of \pc.}
\centering
 \begin{tabular}{ c  c  c  c  c  c  c }
    \hline\hline
  $J^{P}$ & $(1/2)^{+}$   &   $(1/2)^{-}$   &   $(3/2)^{+}$   &   $(3/2)^{-}$ & $(5/2)^{+}$ & $(5/2)^{-}$ \\
   \hline
  Allowed $l$ &  1 & 0 & 1, 3  &  0, 2 & 1, 3 & 2, 4 \\\hline\hline
\end{tabular}
\label{table:roam}
\end{table}

\begin{table*}[htbp]
\caption{Expected \pcftot~yields to be measured by the ePIC experiment at \sqrts=141 GeV ($N_{141}$) and \sqrts=40 GeV ($N_{40}$), assuming an integrated luminosity of 100 fb$^{-1}$. }
\centering
 \begin{tabular}{ c  c  c  c  c  c  c }
    \hline\hline
  $J^{P}$ & $(1/2)^{+}$   &   $(1/2)^{-}$   &   $(3/2)^{+}$   &   $(3/2)^{-}$ & $(5/2)^{+}$ & $(5/2)^{-}$ \\
   \hline $N_{141}$ & 790 & 350 & $5.1\times10^3$ & $2.0\times10^3$ & $4.1\times10^7$ & $1.0\times10^8$\\\hline 
   $N_{40}$ &  $1.7\times10^5$ & $7.7\times10^4$ & $2.8\times10^5$ & $1.1\times10^5$ & $4.8\times10^6$ & $1.2\times10^7$\\
   \hline\hline
\end{tabular}
\label{table:pcyieldamount}
\end{table*}

We also investigated the result at $\sqrt{s} = 40$ GeV, the lowest energy achievable by the EIC in high-luminosity mode~\cite{osti_1670680}. 
At this lower energy, compared to $\sqrt{s} = 141$ GeV, we found that the yield of $P_c(4312)$ is smaller if the spin is $\frac{5}{2}$, while it is significantly larger if the spin is $\frac{1}{2}$ or $\frac{3}{2}$, thereby providing a clear distinction between quantum numbers.  To quantify this behavior, we computed the ratio of the integrated yields at $\sqrt{s} = 141$ GeV and $\sqrt{s} = 40$ GeV, denoted as $N_{141}/N_{40}$, as shown in Figure~\ref{fig:pcamount}.  We considered only the cross section within the pseudorapidity range $|\eta| < 4$ to account for the coverage of the ePIC detector.
The number of $P_c(4312)$ events to be measured by the ePIC experiment can be estimated using the peak instantaneous luminosity of $10^{34} \text{cm}^{-2}\text{s}^{-1}$. Operating at this rate for one year, eight hours per day, would attain an integrated luminosity of 100 fb$^{-1}$ per year. Table~\ref{table:pcyieldamount} shows the expected number of yields under this assumption for different spin-parity scenarios at $\sqrt{s} = 141$ GeV and $\sqrt{s} = 40$ GeV. 
Depending on the spin values, either higher or lower energy provides a statistical advantage. In either case, we expect to collect at least $O(10^4)$ $P_c(4312)$ events, which is statistically ample for precision measurements. 
Furthermore, using the same reasoning, we can confidently anticipate the observation of other heavy-flavor pentaquarks, with the potential for discovering $P_b$ states ($uudb\bar{b}$).

\begin{figure}[H]
\centering
\subfigure[]
{\includegraphics[scale=0.162]{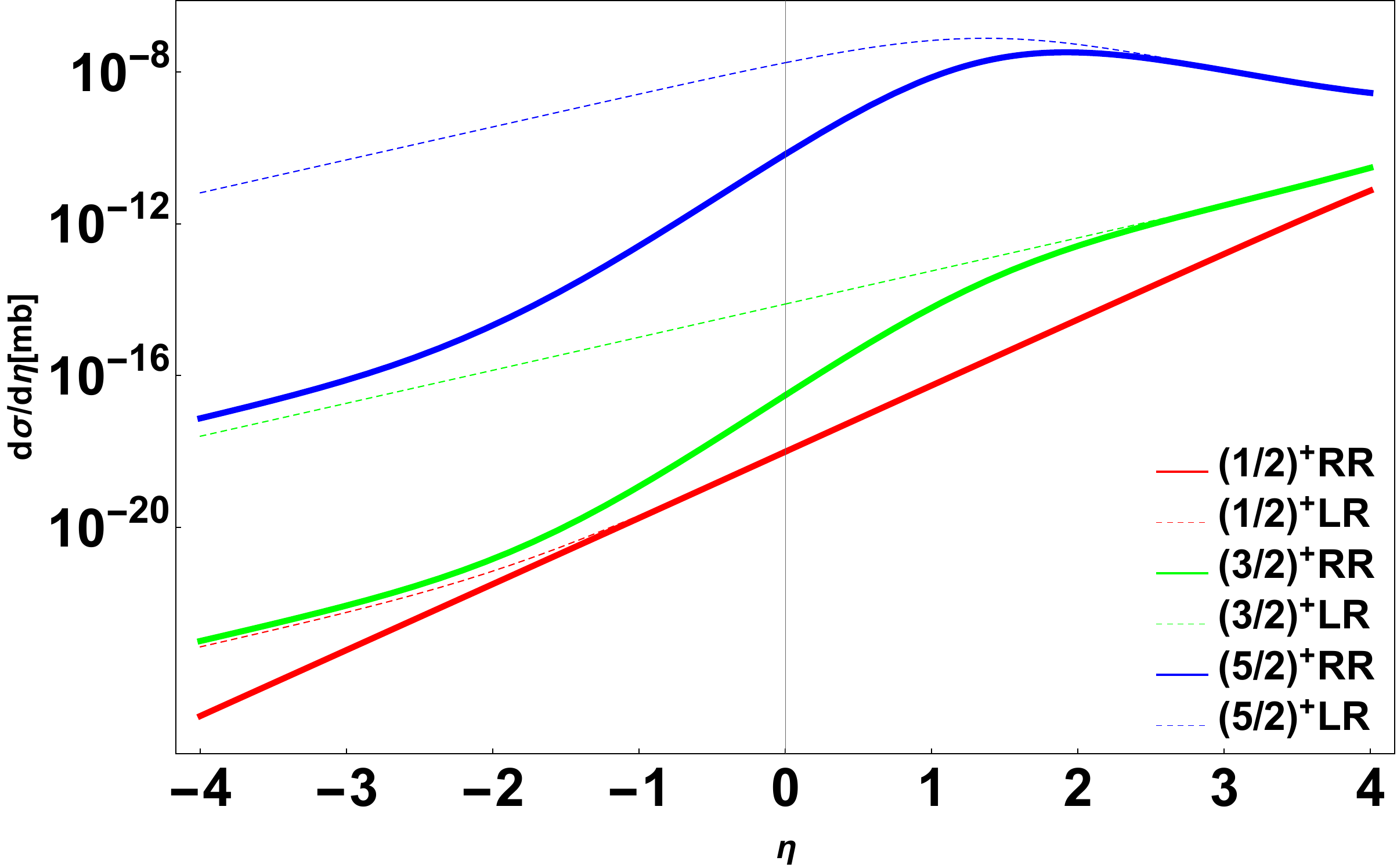}\label{fig:polarized_eta_positiveparity}}
\subfigure[]{\includegraphics[scale=0.20]{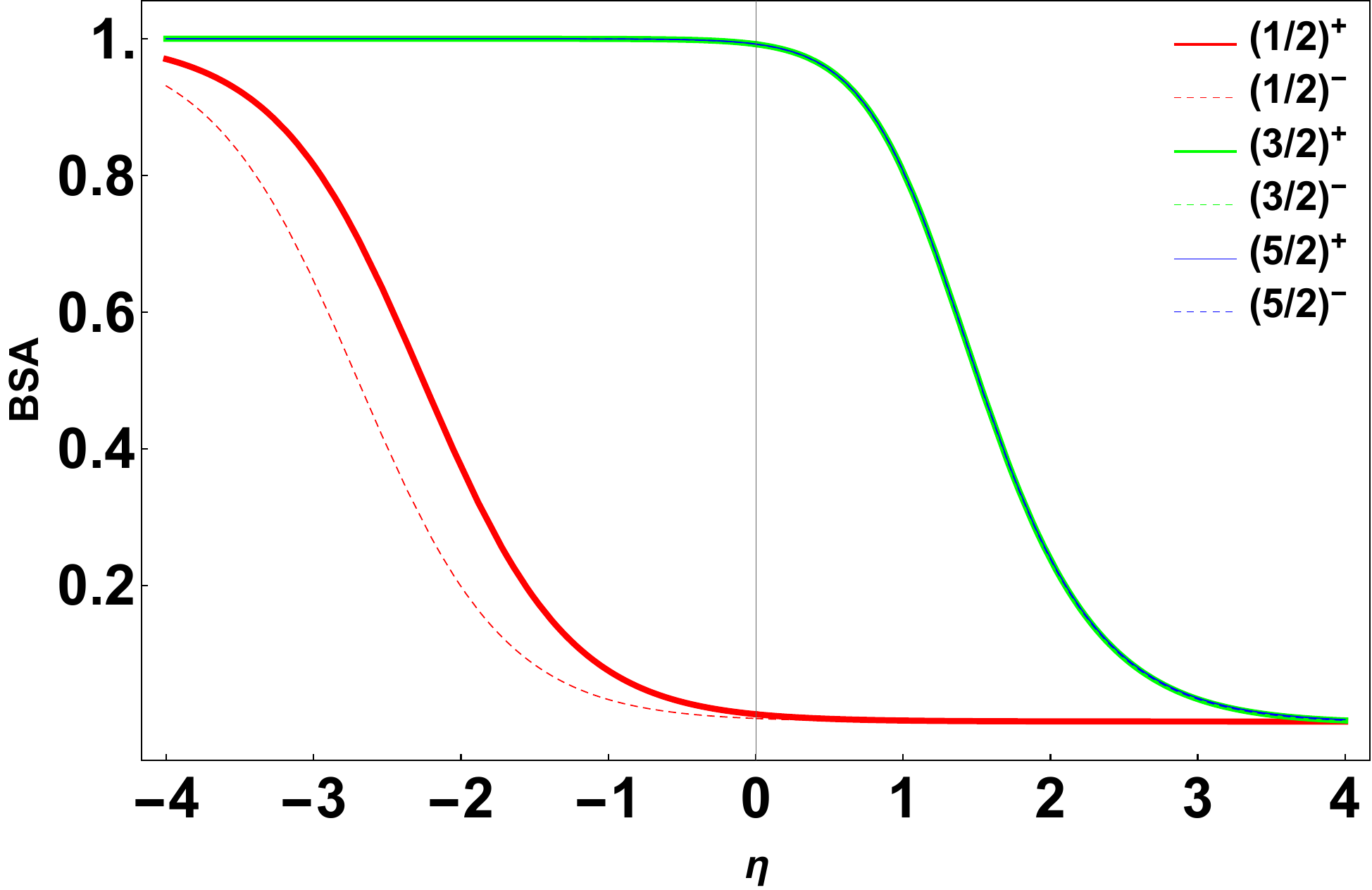}\label{fig:bsa}
}
\caption{(a) Cross sections for different beam polarization configurations under three spin assignments. Only the positive-parity cases are shown in this figure, but the pattern is similar for negative-parity states. 
(b) Beam spin asymmetry (BSA) results for $J^{P}= (1/2)^{\pm},\;(3/2)^{\pm}$ and $(5/2)^{\pm}$ states.  The results for $(5/2)^{\pm}$ and $(3/2)^{\pm}$ are overlapping to each other.}
\end{figure}

{\it Polarization effects:} An additional key advantage of the EIC is its capability to polarize both electron and proton beams. This feature provides a significant benefit in characterizing the spin of pentaquarks, as particles with different spins exhibit distinct production patterns in polarized collisions.  
We denote a right-handed electron colliding with a right-handed proton as \textit{RR} and a right-handed electron colliding with a left-handed proton as \textit{RL}.
The cross sections for \textit{LL} and \textit{LR} are identical to those of \textit{RR} and \textit{RL}, respectively, as only electromagnetic and strong interactions are considered.
The computation is performed by using the projection operator, $P_{\text{R/L}}=(1\pm\gamma_5\slashed{s})/2$, where $s^{\mu}=(0,\bm{s})$ is the spin four-vector in its rest frame.  The result for the $\eta$ dependence is shown in Figure~\ref{fig:polarized_eta_positiveparity}.  
Across all spin scenarios, the \pcftot~cross section in \textit{RL} exceeds that in \textit{RR}.  The deviation is most pronounced in the backward direction and gradually decreases as they align in the forward direction. This behavior can be quantified using the beam spin asymmetry (\textit{BSA}), a standard measure in the spin physics community:
\begin{align}
\label{eq:rfb_bsa}
&BSA~(\eta) = \frac{d\sigma/d\eta~[RL] - d\sigma/d\eta~[RR]}{d\sigma/d\eta~[RL] + d\sigma/d\eta~[RR]} 
\end{align}
This self-normalized observable enhance the experimental precision by eliminating uncertainties related to luminosity, tracking corrections, and geometric acceptance.  The result in Figure~\ref{fig:bsa} shows that the \textit{BSA} can effectively distinguish the spin-$\frac{1}{2}$ case from others.  The spin-$\frac{3}{2}$ and spin-$\frac{5}{2}$ cases  can also be resolved by identifying the different shapes of the differential cross section. As shown in Figure~\ref{fig:polarized_eta_positiveparity},the spin-$\frac{5}{2}$ curve has a maximum cross section near $\eta$ = 2, while the spin-$\frac{3}{2}$ curve monotonically increases within the detector coverage.  

\begin{figure}[H]
\includegraphics[scale=0.14]{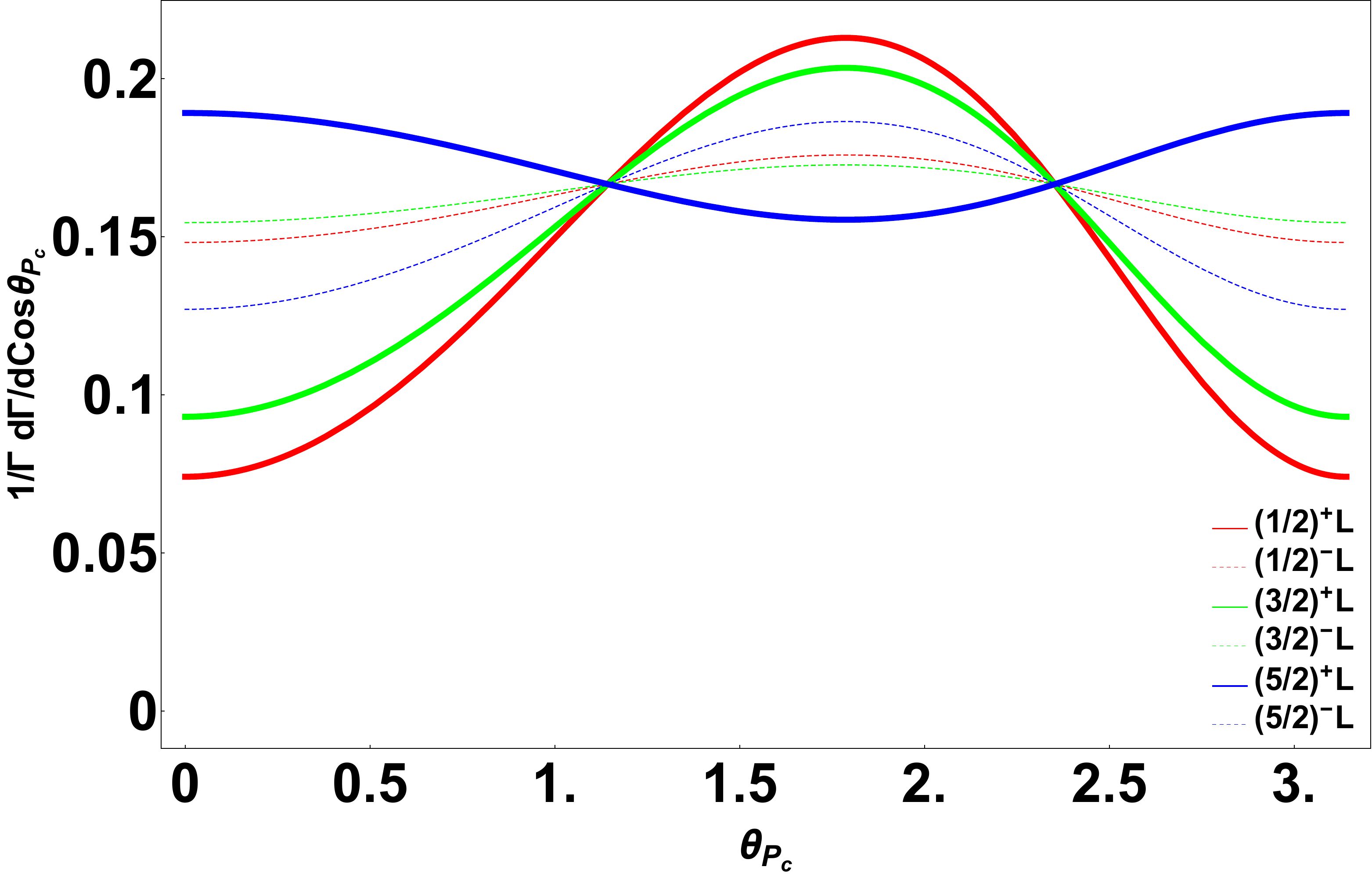}
\caption{ Distribution of  $\theta_{P_c}$ for six spin-parity assignements for \pcftot~, derived from events with longitudinally polarized \jpsi. Refer to the main text for the definition of $\theta_{P_c}$.}\label{fig:jpsipolarization}
\end{figure}

In addition to determining the spin, the parity can also be inferred by analyzing the decay kinematics of its ~\pc~$\rightarrow p + J/\psi \rightarrow p + e^+ + e^- $ channel. The observable is the probability distribution of $\theta_{P_c}$, defined as the polar angle of the \jpsi~decayed from \pc. This angle is measured in the rest frame of \pc, with the z-axis aligned to its momentum in the boosted frame. 
Figure~\ref{fig:jpsipolarization} shows the results for positive and negative parity scenarios of \pc~when longitudinally polarized \jpsi~are used. 
Such longitudinally polarized \jpsi~can be experimentally tagged by analyzing $\theta_{J/\psi}$, defined as the decay angle of the electron from \jpsi, in a manner analogous to $\theta_{P_c}$.  
Given the known distribution of $\theta_{J/\psi}$ one can statistically subtract the contribution from transversely polarized \jpsi~\cite{Park:2022nza}.  To achieve this measurement, which highlights the importance of the detector's hermetic geometry in the EIC experiment, the detector efficiency must be both high and uniform.

{\it Conclusion:} We have demonstrated that the EIC is a powerful facility for both the electroproduction and the characterization of heavy pentaquarks, exemplified by \pcftot.
Its broad collision energy range and high luminosity provide abundant statistics for precise measurements of \pcftot, and the spin-polarized beams enable accurate determination of the spin. 
Through full measurement of decay products, the ePIC experiment can directly reconstruct the invariant mass of \pcftot, offering a significant advantage over GlueX's indirect measurement via inclusive di-electron cross sections.
This approach can be extended to other heavy pentaquarks containing \textit{uud} quarks, such as $P_c(4440)$ and $P_c(4457)$, as they can be electroproduced on a proton. Another unique opportunity at the EIC is the potential discovery of the hypothesized $P_b$ states, composed of $uudb\bar{b}$. Moreover, a distinct class of pentaquark states, potentially including $udd$, can be generated through electron-deuteron collisions at the EIC. Last but not least, a particularly compelling research direction involves studying pentaquark interactions with nuclear matter produced in $e+$A collisions.

Consequently, the EIC stands as a unique pentaquark factory, presenting an unparalleled avenue for a deeper understanding of exotic particles. We strongly encourage the EIC community to fully embrace this potential and leverage its capacity for in-depth pentaquark studies.

\section*{Acknowledgments}
This work was supported by the National Research Foundation of Korea(NRF) grant
funded by the Korea government(MSIT) No. 2022R1A2C1011549 and No. 2018R1A5A1025563 (Y.K.), 
RS-2023-00280831 (S.C.), and 2023R1A2C300302312, 
2023K2A9A1A0609492411 (S.H.L.).  This work is also supported by U.S. Department of Energy under Award DE-SC0012704 (Y.K.).

\clearpage
\appendix

\pagebreak

\end{document}